\pdfoutput=1

\documentclass[11pt,twoside,a4paper,cmspaper,final,collab]{cms-tdr}
\def\svnVersion{167114}\def\cmsCernNo{2012-143}\def\cmsCernDate{\today}\def\cmsMessage{Submitted to the Journal of High Energy Physics}

\begin{document}\cmsNoteHeader{HIN-11-004}

\hyphenation{had-ron-i-za-tion}
\hyphenation{cal-or-i-me-ter}
\hyphenation{de-vices}

\RCS$Revision: 145228 $
\RCS$HeadURL: svn+ssh://svn.cern.ch/reps/tdr2/papers/HIN-11-004/trunk/HIN-11-004.tex $
\RCS$Id: HIN-11-004.tex 145228 2012-08-29 12:29:54Z cer $
\providecommand {\mubinv}{\ensuremath{\mathrm{\mu b^{-1}}}}
\newcommand{\sNN}{\ensuremath{\sqrt{s_{_{\mathrm{NN}}}}}}
\newcommand {\roots}    {\ensuremath{\sqrt{s}}}
\newcommand {\rootsNN}  {\ensuremath{\sqrt{s_{_{NN}}}}}
\newcommand {\dndy}     {\ensuremath{dN/dy}}
\newcommand {\dnchdy}   {\ensuremath{dN_{\mathrm{ch}}/dy}}
\newcommand {\dndeta}   {\ensuremath{dN/d\eta}}
\newcommand {\dnchdeta} {\ensuremath{dN_{\mathrm{ch}}/d\eta}}
\newcommand {\dndpt}    {\ensuremath{dN/d\pt}}
\newcommand {\dnchdpt}  {\ensuremath{dN_{\mathrm{ch}}/d\pt}}
\newcommand {\deta}     {\ensuremath{\Delta\eta}}
\newcommand {\dphi}     {\ensuremath{\Delta\phi}}
\newcommand {\dNdxsi} {\ensuremath{dN_{\mathrm{track}}/d\xi}}

\newcommand {\AJ}       {\ensuremath{A_J}}
\newcommand {\npart}    {\ensuremath{N_{\rm part}}}
\newcommand {\ncoll}    {\ensuremath{N_{\rm coll}}}

\newcommand {\pp}    {\mbox{pp}}
\newcommand {\ppbar} {\mbox{p\={p}}}
\newcommand {\pbarp} {\mbox{p\={p}}}
\newcommand {\PbPb}  {\mbox{PbPb}}

\newcommand{\m}{\ensuremath{\,\text{m}}\xspace}

\newcommand {\naive}    {na\"{\i}ve}
\providecommand{\PHOJET} {\textsc{phojet}\xspace}

\def\d{\mathrm{d}}

\providecommand{\PKzS}{\ensuremath{\mathrm{K^0_S}}}
\providecommand{\Pp}{\ensuremath{\mathrm{p}}}
\providecommand{\Pap}{\ensuremath{\mathrm{\overline{p}}}}
\providecommand{\PgL}{\ensuremath{\Lambda}}
\providecommand{\PagL}{\ensuremath{\overline{\Lambda}}}
\providecommand{\PgS}{\ensuremath{\Sigma}}
\providecommand{\PgSm}{\ensuremath{\Sigma^-}}
\providecommand{\PgSp}{\ensuremath{\Sigma^+}}
\providecommand{\PagSm}{\ensuremath{\overline{\Sigma}^-}}
\providecommand{\PagSp}{\ensuremath{\overline{\Sigma}^+}}

\providecommand{\HYDJET}{\textsc{hydjet}\xspace}
\cmsNoteHeader{HIN-11-004} 
\title{Measurement of jet fragmentation into charged particles
in pp and PbPb collisions at $\sqrt{s_{\mathrm{NN}}}= 2.76\TeV$ }

\date{\today}

\abstract{
Jet fragmentation in pp and PbPb collisions at a centre-of-mass energy
of 2.76 TeV per nucleon pair was studied using data collected with the CMS
detector at the LHC.
Fragmentation functions are constructed using charged-particle tracks with transverse momenta $\pt > 4\GeVc$ for dijet events with a leading jet of $\pt > 100\GeVc$.
The fragmentation functions in PbPb events are compared to those in pp data as a function of collision centrality, as well as dijet-$\pt$ imbalance.
Special emphasis is placed on the most central PbPb events including dijets with
unbalanced momentum, indicative of energy loss of the hard scattered parent partons.
The fragmentation patterns for both the leading and subleading jets in PbPb
collisions agree with those seen in pp data at 2.76\TeV.
The results provide evidence that, despite the large parton energy loss
observed in PbPb collisions, the partition of the remaining momentum within 
the jet cone into high-\pt particles is not
strongly modified in comparison to that observed for jets in vacuum.
}

\hypersetup{%
pdfauthor={CMS Collaboration},%
pdftitle={Measurement of jet fragmentation into charged particles in pp and PbPb collisions at sqrt(s[NN])= 2.76 TeV},%
pdfsubject={CMS},%
pdfkeywords={CMS, physics, heavy ions, jet quenching}}

\maketitle 

\section{Introduction}

When colliding lead nuclei (PbPb) at a nucleon-nucleon centre-of-mass energy
of $\sNN=2.76\TeV$, one expects to form a system of hot
and dense matter at energy densities that have not been explored before.
One of the early proposed experimental signatures of the formation of a dense system in such
collisions was
``jet quenching'', \ie the suppression or disappearance
of the spray of hadrons resulting from the fragmentation of a hard scattered
parton having suffered energy loss in the medium~\cite{Bjorken:1982tu}.
The energy lost by a parton in the produced medium provides fundamental
information on its thermodynamical and transport
properties~\cite{CasalderreySolana:2007zz,d'Enterria:2009am}.
Results on the suppression of inclusive hadron production at
high transverse momenta (\pt), as well as on the modified high-\pt dihadron
angular correlations obtained from nucleus-nucleus collisions
at $\sNN =200\GeV$ at
the Relativistic Heavy Ion Collider
(RHIC)~\cite{Arsene:2004fa,Back:2004je,Adams:2005dq,Adcox:2004mh}
have shown the existence of partonic energy loss in dense strongly
interacting matter. Similar observations have been made at the
Large Hadron Collider (LHC)~\cite{Aamodt:2010jd,CMS-HIN-10-005,Aad:2012bu,Chatrchyan:2012xq,Aamodt:2010pa,Chatrchyan:2012ta,ATLAS:2011ah}.

At LHC energies, high-\pt\ jets have been fully reconstructed in heavy-ion
collisions. A significant dijet transverse momentum imbalance is observed,
when comparing to a reference distribution corresponding to pp collisions at the same
nucleon-nucleon centre-of-mass energy~\cite{Collaboration:2010bu,Chatrchyan:2011sx,CMS-HIN-11-013}.
Such an observation is
consistent with energy loss of the hard scattered partons in the dense medium produced in
central \PbPb\ collisions.
In the same set of results, the redistribution of the lost jet energy is studied
using jet-track correlations~\cite{Chatrchyan:2011sx}. It is found that the
missing \pt opposite to the leading jet can be recovered only by summing up the
contributions of particles down to a low \pt of $500\MeVc$ with respect to the
beam axis and out to large radii in pseudorapidity and azimuthal angle, $\Delta R=\sqrt{(\Delta\phi)^2+(\Delta \eta)^2} > 0.8$, with respect to the jet axis~\cite{Chatrchyan:2011sx}.
Since these results show that the quenched
energy is transferred out of the jet cone, the jet clustering algorithm reconstructs
jets with a reduced energy from the fragments of the partons after they have lost
energy in the medium.
The study presented here investigates to what extent the fragmentation
pattern of
partons that have traversed the medium resembles vacuum fragmentation, by constructing
fragmentation functions in \PbPb\ collisions and comparing them to those from
unquenched jets, as measured in pp collisions.
Fragmentation functions encode the probability for a parton to fragment into
particles carrying a given fraction of the parton energy. Colour-charged partons
undergo showering processes into partons of successively lower energy which
hadronise into colour-neutral final-state particles.
The evolution of such a parton radiation and splitting process leads to a
characteristic shape of the fragmentation function \cite{Dokshitzer:1991wu}.
Theoretical models
of jet quenching predict an effective change of the shape of the fragmentation
function due to the change of the parton radiation pattern in the medium~
\cite{Guo:2000nz,Borghini:2005em,Armesto:2007dt,Arleo:2008dn}.
Experimentally, fragmentation functions are constructed by correlating the
reconstructed jet momentum with the momenta of charged particles projected onto
the jet axis.
The jets are defined using the final-state particles produced in the collision, clustered with the anti-$k_\mathrm{T}$ jet algorithm~\cite{Cacciari:2008gp}.
In this Letter, we present a measurement of fragmentation
functions in \pp\ and \PbPb\ collisions and a detailed comparison of their shapes measured in the two systems.
The measurement is restricted to the
high-\pt component of the fragmentation function, using charged particles
of $\pt>4\GeVc$ that lie within $\Delta R < 0.3$ around the reconstructed jets.

\section{Experimental Setup}
\label{sec:results}

The Compact Muon Solenoid (CMS) detector is described in Ref.~\cite{bib_CMS}. Only the
detector systems used in this analysis are discussed hereafter. The
central part of the CMS detector contains a superconducting solenoid
that provides a homogeneous magnetic field of 3.8\unit{T} parallel to the beam
axis. Charged-particle trajectories are measured using silicon pixel
and strip trackers that cover the pseudorapidity region $|\eta| <
2.5$, where $\eta$ is defined as $\eta = -\log\left[\tan\left(\theta/2\right)\right]$ and
$\theta$ is the polar angle with respect to the anticlockwise beam direction.
An electromagnetic crystal calorimeter (ECAL) and a
brass/scintillator hadron calorimeter (HCAL) surround the tracking
volume and cover $|\eta|< 3.0$.
The ECAL calorimeter is segmented in quasi-projective cells of a granularity
in pseudorapidity and azimuthal angle of $\Delta \eta \times
\Delta\phi=0.0174\times0.0174$ in the barrel ($|\eta| <1.5$), increasing
across the endcap ($1.5<|\eta|<3.0$) to  $0.09 \times 0.09$ at $|\eta| = 3.0$.
The HCAL has a cell granularity of $\Delta \eta \times \Delta\phi
=0.087\times0.087$ in the barrel region and
 a variable cell granularity, changing as a function of $\eta$,
in the endcap region~\cite{bib_CMS}.
A forward steel/quartz-fibre Cherenkov hadron calorimeter (HF) extends
the coverage to $|\eta| = 5.2$.
The CMS trigger system is composed of a first level made of custom
hardware processors, which use information from the calorimeters and muon
detectors to select events, and the High-Level Trigger (HLT) processor
farm, that further decreases the event rate, before data storage.

\section{Data Selection}

The PbPb and pp data analysed in this Letter were collected with the CMS
detector in 2010 and 2011, respectively, at a centre-of-mass energy of 2.76\TeV per nucleon pair.
The integrated luminosities for the PbPb and pp data samples used for this analysis
are $\mathrm{L}_\text{int}\approx6.8\mubinv$ and $\mathrm{L}_\text{int}\approx231\nbinv$, respectively.
The HLT system is used to select events containing
high-\pt\ jets reconstructed from calorimeter towers.
In PbPb collisions, the trigger
threshold is $\pt = 35\GeVc$, applied on the raw calorimetric jet energy.
For pp collisions, events are selected if they pass a jet trigger
threshold of $\pt = 40\GeVc$ on the calorimetric jet energy.
As found in Ref.~\cite{Chatrchyan:2011sx}, for the jet selection used in this
analysis, requiring a 100 \GeVc jet in $|\eta| < 2$, the triggers are more than 99\% efficient.
In addition to the trigger decision, standard event selection criteria are applied ~\cite{Chatrchyan:2011sx}, including a rejection of beam related backgrounds,
a selection of inelastic hadronic collisions by requiring a two-sided coincidence of signals in the HF and a well-reconstructed event vertex.

For the analysis of PbPb data, it is important to determine the degree of
overlap between the two colliding nuclei in each event, termed collision centrality.
Centrality is determined using the sum of transverse energy reconstructed
in the HF. The distribution of the HF energy is used to divide the
event sample into percentiles of the total nucleus-nucleus interaction cross section.
For the purpose of this analysis, the data set is split into two centrality bins,
the 0--30\% most central events
(\ie those which have the largest overlap between the two colliding Pb nuclei)
and the remaining peripheral events in
the 30--100\% centrality range. A detailed description of the centrality
determination can be found in~\cite{Chatrchyan:2011sx}.

\section{Jet and Track Reconstruction}

For both \pp\ and \PbPb\ collisions, the analysis is based on jets reconstructed using
the anti-$k_\mathrm{T}$ jet algorithm, with a radius parameter ($R$)
of 0.3, utilizing particle-flow (PF) objects that combine
tracking and calorimetric information~\cite{MattPFlow,particle_flow}.
In the \PbPb\ data, the contribution of the underlying heavy-ion event is
removed using an iterative pileup subtraction method~\cite{Kodolova:2007hd}.
Since this procedure determines the underlying-event background using data  
outside the jet, the result is insensitive to the fragmentation properties of the jet.
The jet-finding efficiency is above 95\% for jets of \pt $>40~\GeVc$, and above 99\% for
jets of $\pt >50\GeVc$~\cite{MattPFlow}.
The relative jet momentum resolution in pp collisions is found to be 19~(13)\% at $\pt =40\ (100)\GeVc$, improving with jet momentum. In central PbPb collisions, the momentum resolution deteriorates to 24~(16)\% at $\pt =40\ (100)\GeVc$~\cite{MattPFlow} due to fluctuations in the underlying event. For both pp and PbPb events,
the jet momentum response has little or no deviations from a Gaussian shape.

In all cases the reconstructed jet momenta are corrected to final-state
stable particle (lifetime $\tau$ with $c\tau>10\unit{mm}$) level using factors derived
from \PYTHIA~6.422~\cite{bib_pythia} (tune D6T~\cite{TUNE-D6T-1,TUNE-D6T-2},
CTEQ6L1 PDFs~\cite{CTEQ6L1}) pp simulations at $\sqrt{s} = 2.76\TeV$~\cite{Chatrchyan:2011ds}.
The uncertainty in the corrected jet energy scale is about 3\% for pp events, resulting in a per-bin jet-yield uncertainty of
${\pm}15\%$.
In the case of PbPb events, due to the influence of the underlying event, the
uncertainty in the jet energy scale increases to about 4\% for peripheral
events (30--100\% centrality) and 5\% for central events (0--30\% centrality) which results in per-bin jet-yield uncertainties of ${\pm}20\%$ and ${\pm}25\%$, respectively.

The dijets selected for this analysis consist of a leading jet (denoted by subscript 1)
with $p_{\mathrm{T},1} > 100$\GeVc and a subleading jet (subscript 2)
of $p_{\mathrm{T},2} > 40$\GeVc, with axes that lie within $|\eta| < 2$.
The \pt\ thresholds are chosen to ensure high reconstruction efficiency for
the leading and, especially, the subleading jet.
In addition, the azimuthal opening angle $\Delta \phi_{1,2}$
between the leading and subleading jet is required to be larger than $2\pi/3$. No explicit requirement is made on the presence or absence of
a third jet in the event.

A detailed description of the charged-particle reconstruction algorithm and its performance
can be found in Ref.~\cite{CMS-HIN-10-005}. The track-finding efficiency in the kinematic range
of this study is (60--70)\% and the corresponding correction is applied as a function of
track \pt, jet \pt, and event centrality by reweighting the found tracks with the inverse of the reconstruction efficiency.
The track reconstruction efficiency correction is derived from a \GEANTfour \cite{bib_geant}
simulation of the CMS detector applied to \PYTHIA events, which are embedded into \PbPb\ collisions simulated using \HYDJET~\cite{Lokhtin:2005px} in order to include the effect of the underlying PbPb event.
The momentum resolution of the track reconstruction is $\sigma(\pt)/\pt \approx1$--3\%.

\begin{figure}[htb]
\begin{center}
\includegraphics[width=1.0\textwidth]{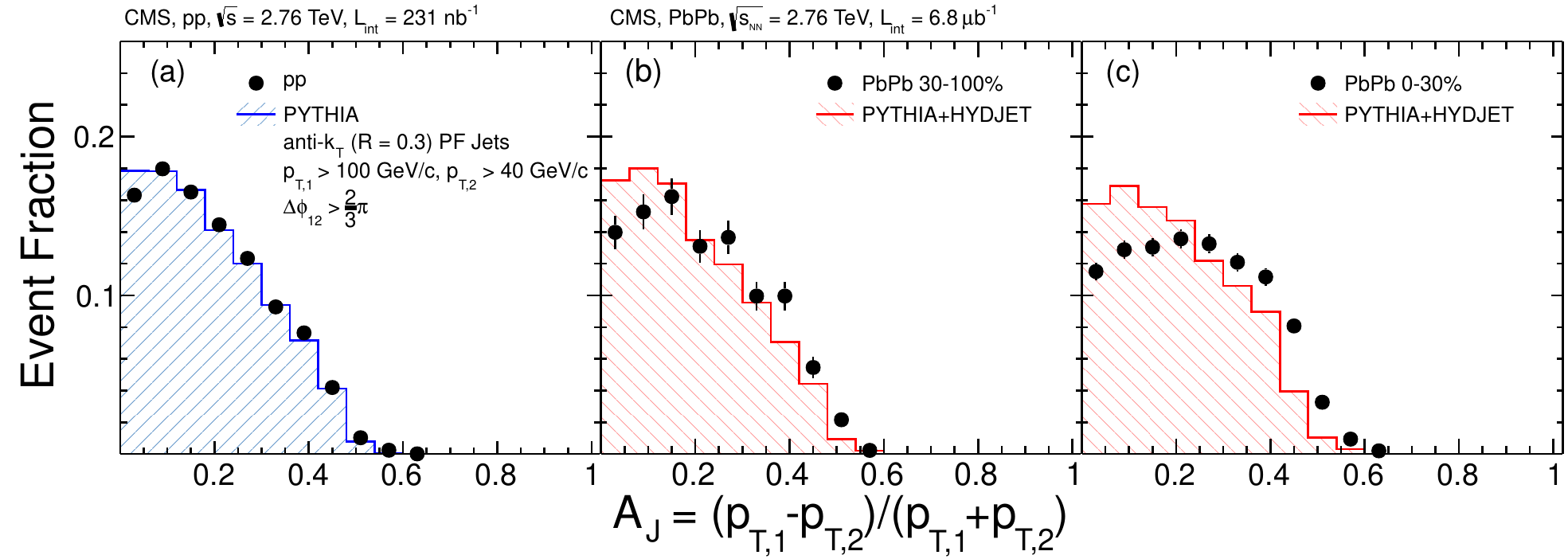}
\end{center}
\caption{Dijet asymmetry, $\AJ$, distributions
in (a) pp collisions, (b) peripheral (30--100\%) PbPb, and (c) central (0--30\%) PbPb collisions. Data are shown as black points while the histograms
show \PYTHIA dijets, which when compared to PbPb data have been embedded
into \HYDJET events.
Error bars represent the statistical uncertainty.}
\label{fig:imbalance}
\end{figure}

The dijet momentum balance is studied in terms of the dijet asymmetry ratio~\cite{Collaboration:2010bu,Chatrchyan:2011sx,CMS-HIN-11-013},
\begin{equation}
\label{eq:aj}
A_J = \frac{p_{\mathrm{T},1}-p_{\mathrm{T},2}}{p_{\mathrm{T},1}+p_{\mathrm{T},2}},~
\end{equation}
which is positive by construction.
Figure \ref{fig:imbalance} shows the $\AJ$ distributions in (a) \pp\ and in (b,c) \PbPb\
for two bins in event centrality.
Central \PbPb\ events (0--30\%) show a significant excess of unbalanced pairs when
compared to both peripheral PbPb collisions (30--100\%) and pp data.
This can be interpreted as a direct observation of parton energy loss
in central \PbPb\ collisions.

\section{Fragmentation Functions}

The fragmentation functions are measured by correlating reconstructed
charged-particle tracks falling within the jet cones, with the axis of the respective jet~\cite{PhysRevD.64.032001}.
As done in previous measurements at hadron colliders \cite{PhysRevLett.65.968},
the fragmentation function is presented as a function of the variable

\begin{equation}
\label{eq:xi}
\xi = \ln \frac{1}{z} ~ ~~;~~  z = \frac{p_{\parallel}^{\text{track}}}{p^{\text{jet}}},
\end{equation}

where $p_{\parallel}^{\text{track}}$ is the momentum component of the track along
the jet axis, and $p_{\text{jet}}$ is the magnitude of the jet momentum within the jet cone.
The momentum components and the angle between the charged-particle and the jet axis
are calculated in the dijet centre-of-mass frame, obtained by an approximate
Lorentz transformation along the beam axis in the form of a pseudorapidity shift,
defined as $\eta_{\text{dijet}} = (\eta_{1}+\eta_{2})/2$.
The tracks are selected to lie within a cone of
$\Delta R < 0.3$
 around the jet axis.
The fragmentation functions, defined as $1/N_\text{jet}$ \dNdxsi, are normalised
to the total number of selected leading or subleading jets, $N_\text{jet}$, respectively.
To minimise the contribution of tracks from the underlying event, only tracks
with $\pt^\text{track} > 4\GeVc$ are selected. This restricts the measurement
 of the fragmentation function to the region of $\xi \lesssim4.5$.
The remaining underlying
event contribution, not associated with the jet, is estimated by selecting tracks
that lie in a background cone, obtained by reflecting the original jet cone
about $\eta$ = 0, while keeping the same $\phi$ coordinate.
The background contribution is accumulated jet-by-jet over the full event sample
and subtracted to obtain the final \dNdxsi\ distribution.
Due to this procedure, jets in the region $|\eta| <$ 0.3 are excluded from the analysis,
to avoid overlap between the signal jet region and the region used for background
estimation.

\begin{figure}[htbp]
\begin{center}
\includegraphics[width=.45\textwidth]{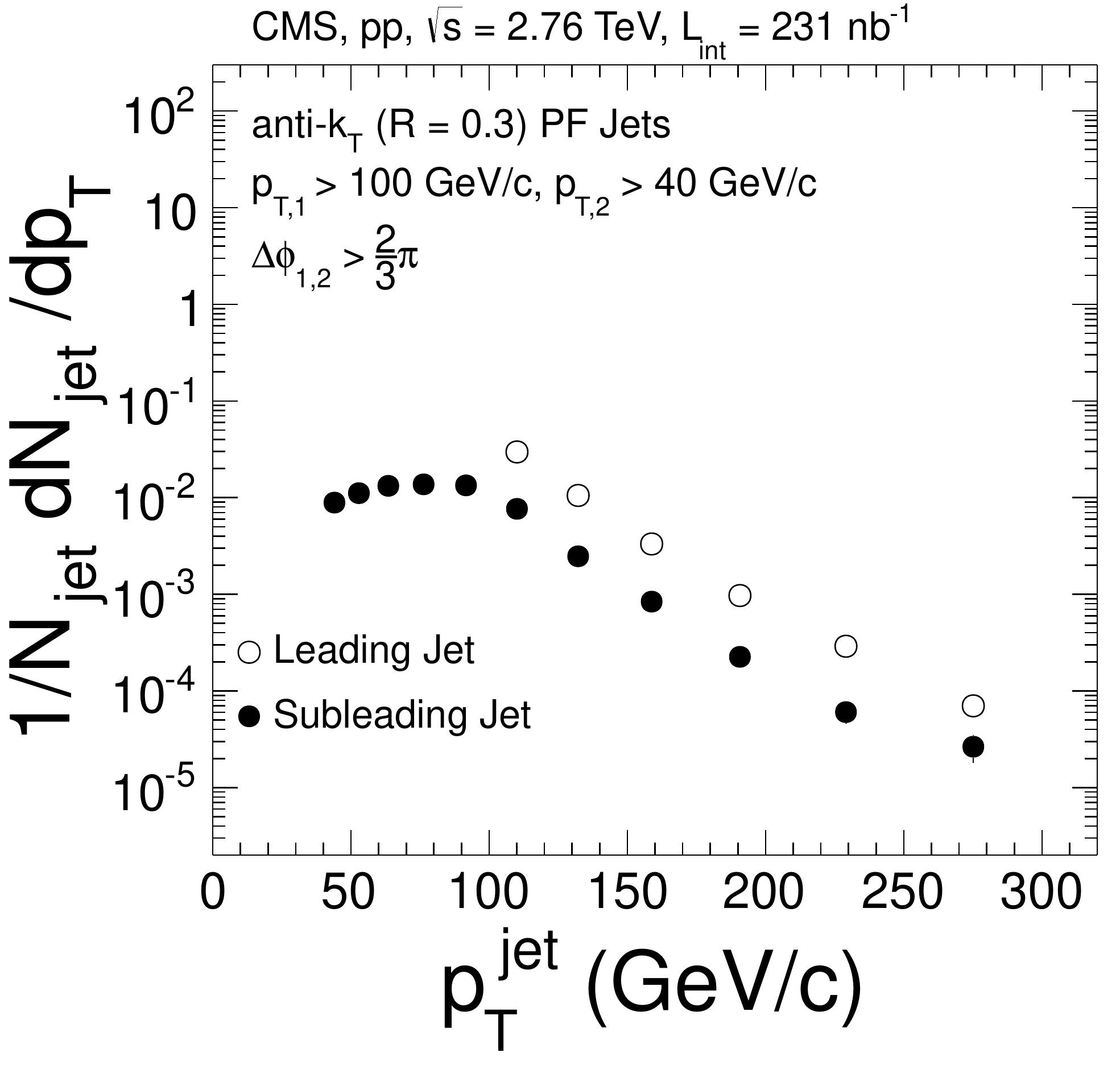}
\includegraphics[width=.45\textwidth]{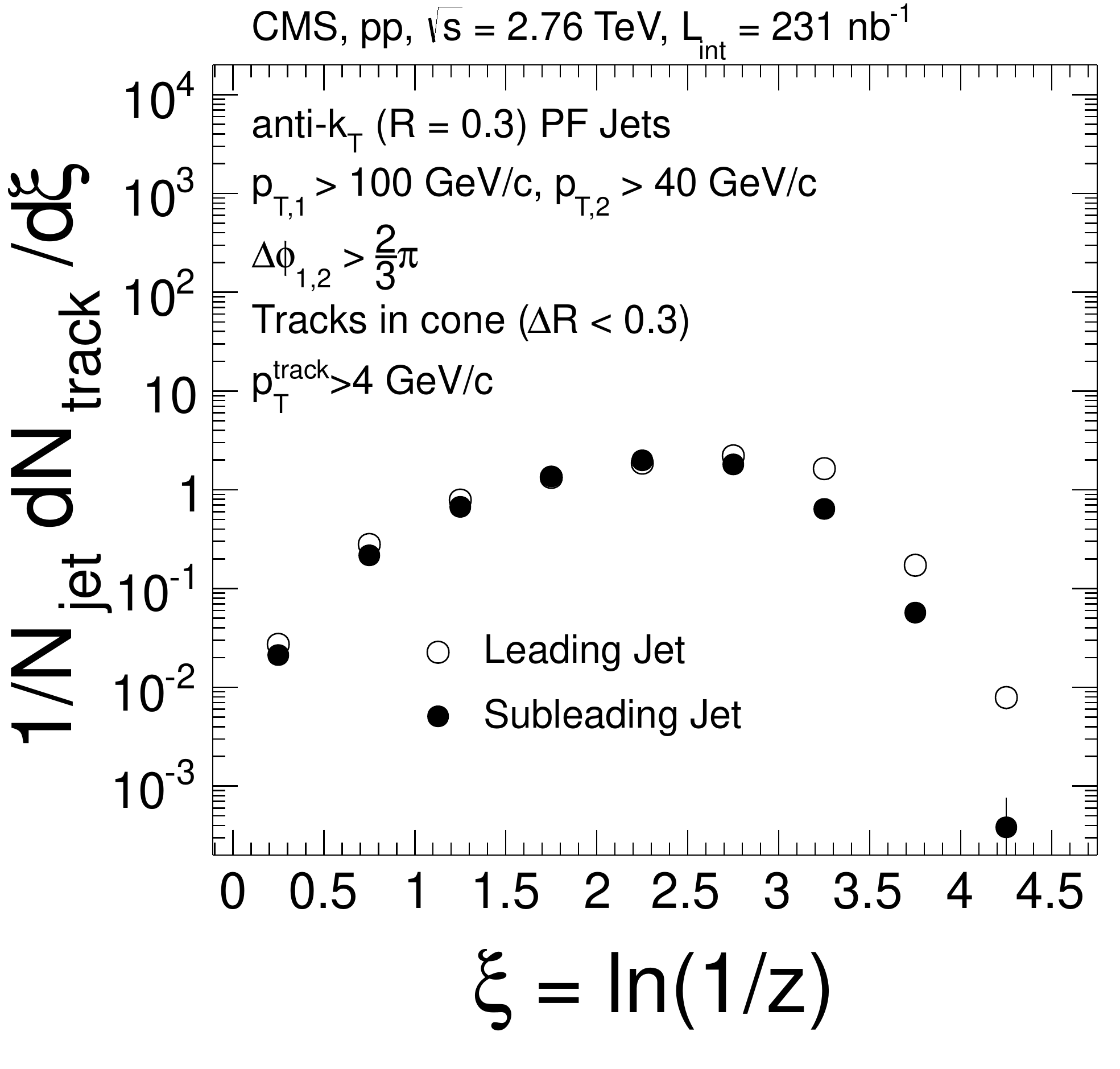}
\end{center}
\caption{
Data from pp collisions.
Left: Leading and subleading jet $\pt$ distributions (not corrected for jet-finding efficiency and not unfolded for the jet energy resolution).
Right: Fragmentation functions reconstructed for the leading (open circles) and subleading (solid points) jets.  The statistical uncertainties, shown as error bars, are smaller than the symbols in most cases.
}
\label{fig:ffLeadSublead_pp}
\end{figure}

\begin{figure}[htbp]
\begin{center}
\includegraphics[width=0.9\textwidth]{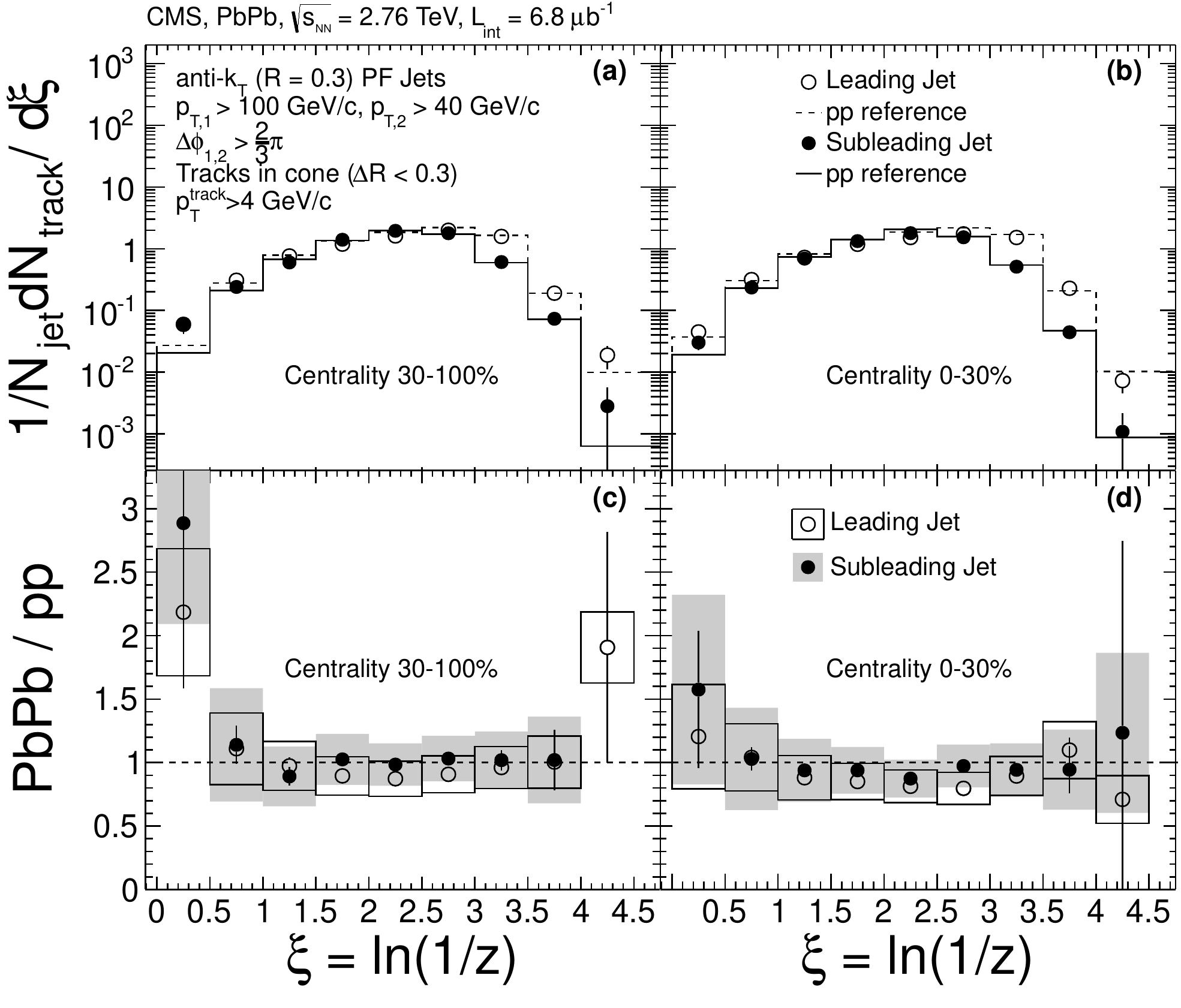}
\includegraphics[width=0.9\textwidth]{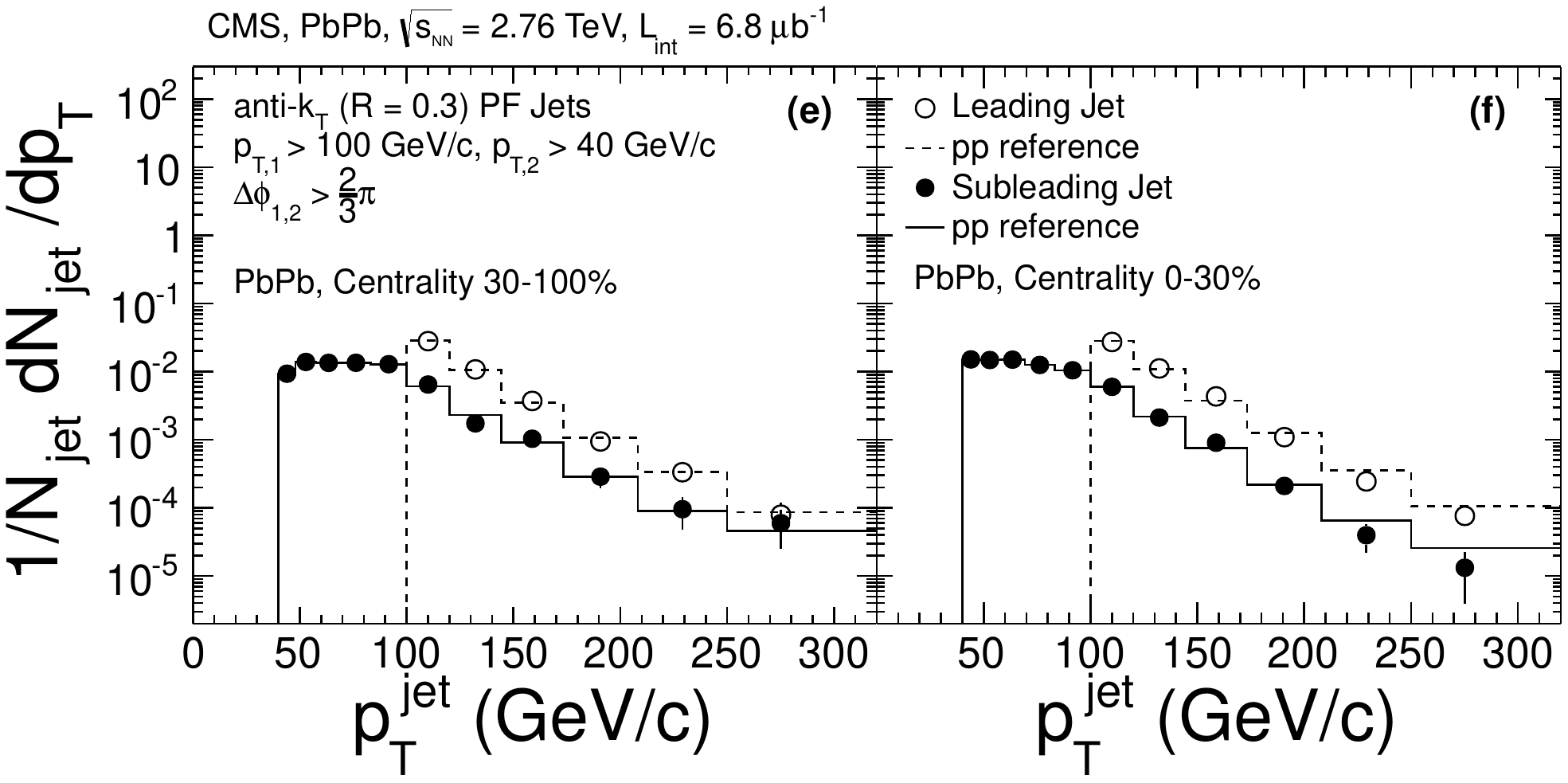}
\end{center}
\caption{(a,b) Fragmentation functions reconstructed in peripheral and central PbPb data for the leading (open circles) and subleading (solid points) jets. (c,d) Ratio of each PbPb fragmentation function to its
pp-based reference.
Error bars are statistical, the hollow boxes represent the systematic uncertainty for the leading jet, and gray boxes show the systematic uncertainty for the subleading jet.
(e,f) Jet \pt\ distributions in PbPb data (not corrected for efficiency and not unfolded for \pt\
resolution) compared to the
pp-based reference
(see text). Only statistical uncertainties are shown in panels a, b, e and f.
}
\label{fig:ffLeadSublead_pp_ppDiv}
\end{figure}

Figure~\ref{fig:ffLeadSublead_pp} shows the reconstructed leading and subleading jet
fragmentation functions in pp collisions (right panel) and the corresponding
jet \pt\ distributions (left panel), to illustrate the kinematic range in which the
fragmentation functions are measured.
Note that the higher jet momentum of the leading jet compared to the subleading jet,
leads to an increased number of particles passing the $\pt^\text{track} > 4\GeVc$ selection
for the fragmentation function measurement.
This results in the observed excess of \dNdxsi\ at high values of $\xi$ for the leading
jet over the corresponding distribution for the subleading jet.
The $\pt^\text{track}$ threshold on the tracks introduces a jet-\pt-dependent kinematic limit in the high $\xi$ part of the spectrum.
For a direct comparison between pp and PbPb collisions, the jet momentum resolution
deterioration in \PbPb\ events has to be taken into account.
For this purpose, the reconstructed \pt\ of every jet in the pp data is
smeared by the quadratic difference of the underlying event fluctuations
in \PbPb\ and \pp.
Furthermore, in order to keep the kinematic constraints consistent, a jet-\pt-dependent
reweighting is applied to the pp data, after fluctuation smearing, so that the resulting jet \pt\ distribution matches that in PbPb.
The reweighting factor is applied to each jet when generating the fragmentation function for pp.
The {\it pp-based reference} distributions obtained this way ensure that the jet fragmentation functions in \PbPb\ and \pp\ are compared for matching jet \pt\ spectra.

Figure~\ref{fig:ffLeadSublead_pp_ppDiv} shows the fragmentation functions for
(a) peripheral and (b) central \PbPb\ collisions, for both the leading and subleading jets, compared to the pp-based reference.
The ratios between the PbPb fragmentation functions and the pp-based reference distributions
are shown in panels (c) and (d).
The corresponding jet \pt\ distributions illustrating the kinematic range of the
measurement are shown in panels (e) and (f).
The overlaid histograms in the same set of figures show the pp-based reference distributions.
The systematic uncertainty, represented by the boxes at each point in
panels (c) and (d), is obtained from the propagated jet and track reconstruction
uncertainties.
These comparison plots show that the shape of the fragmentation
functions in pp and PbPb collisions agrees within uncertainties at all centralities
for the leading, as well as for the subleading, jets.

The uncertainties in the jet response may affect the results in different ways:
smearing of jet energy due to fluctuations distorts the observed fragmentation
functions, a miscalibration of the overall energy scale shifts the fragmentation
function along the $\xi$ axis, and a residual offset in the jet energy introduces
a tilt of the shape of the distribution.
These effects are studied using a Monte Carlo (MC) simulation, by varying the corresponding
 generator-level jet properties within the limits of the jet response uncertainty.
The systematic uncertainties are determined by comparing the  resulting
fragmentation functions in the modified sample to the original MC reference.

The systematic uncertainty due to the charged-particle reconstruction efficiency
is obtained by comparing fragmentation functions based on efficiency-corrected
tracks, with the fragmentation functions using the MC generator information.
Since the particle-flow event reconstruction algorithm uses reconstructed
charged-particles for the jet \pt\ determination, a failure to reconstruct a high-\pt\
charged-particle can lead to an underestimation of the jet momentum, resulting in an artificially
high $\AJ$ measurement. The modification of the fragmentation function measurement due
to this effect is studied in $\PYTHIA+\HYDJET$ events. Since the $\AJ$ distributions
in data and simulation are different, the magnitude of the corresponding effect in
the PbPb data is estimated based on the reconstructed $\AJ$ distributions, and is
accounted for in the combined systematic uncertainty.
The effect of momentum resolution on reconstructed charged-particle tracks is estimated
by smearing $\PYTHIA+\HYDJET$ generator-level information. This is found to have
little effect on the fragmentation function, in comparison to the unsmeared generator
level information. The above uncertainties are combined in quadrature to give the total
systematic uncertainty.

Another potential source of systematic uncertainty comes from the response of the PF technique
to jets with very different fragmentation functions. Since one component of the jet energy
correction accounts for the loss of low-\pt\ particles, a large change in the contribution of such tracks 
will result in an incorrect reconstructed jet energy. To study this effect, we have compared
the response to separate quark and gluon jets from $\PYTHIA$, whose fragmentation functions differ by
20-40\% in the region $2<\xi<4$ and by larger factors for $\xi<1$. These dramatic differences in
fragmentation pattern do result in systematic differences in the jet momentum but the effect is
only at the few percent level \cite{MattPFlow}. The change in the reconstructed fragmentation function
resulting from the tiny shifts in the $\xi$ parameter caused by these jet momentum offsets are negligible.

\begin{figure}[htb]
\begin{center}
\includegraphics[width=1.0\textwidth]{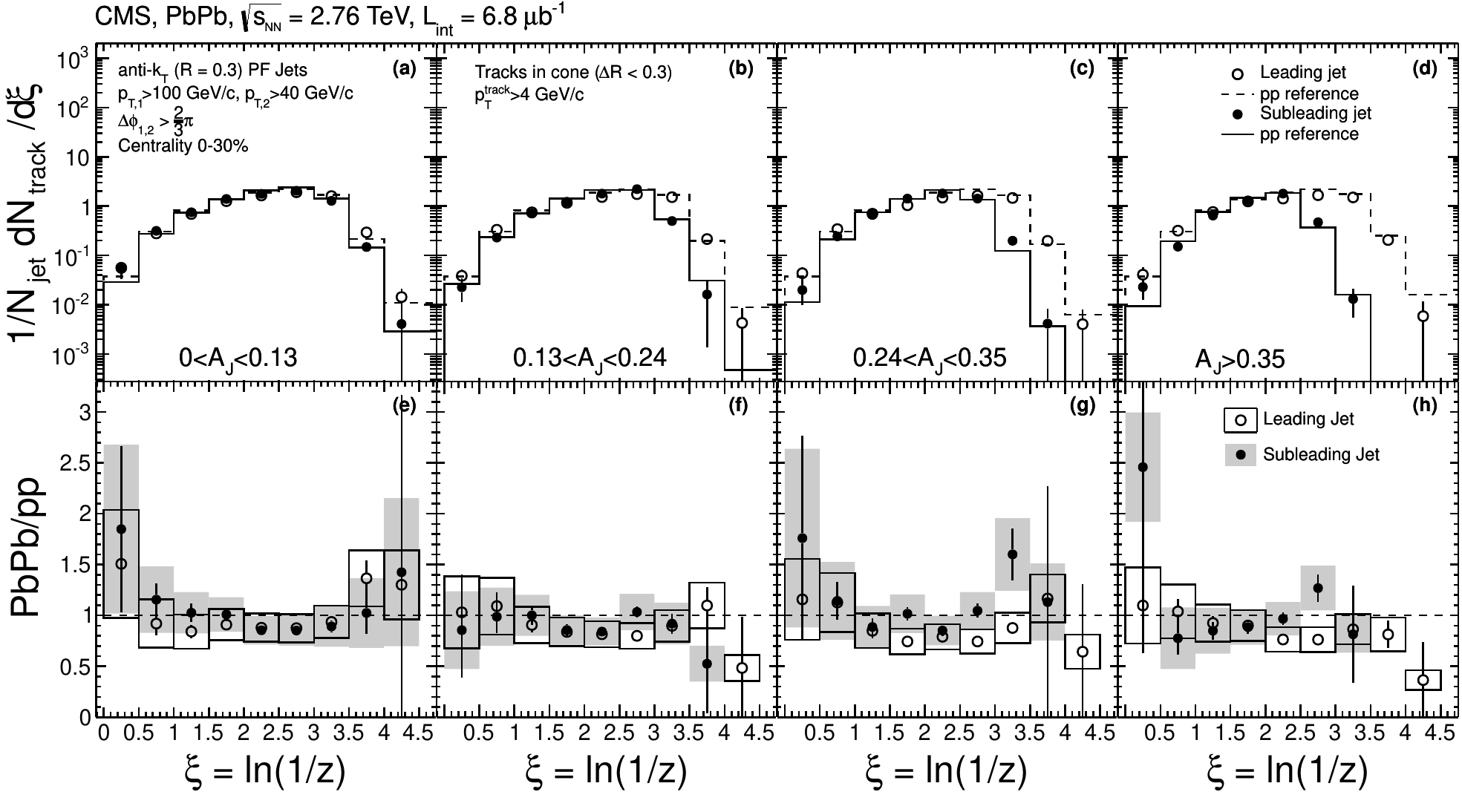}
\includegraphics[width=1.0\textwidth]{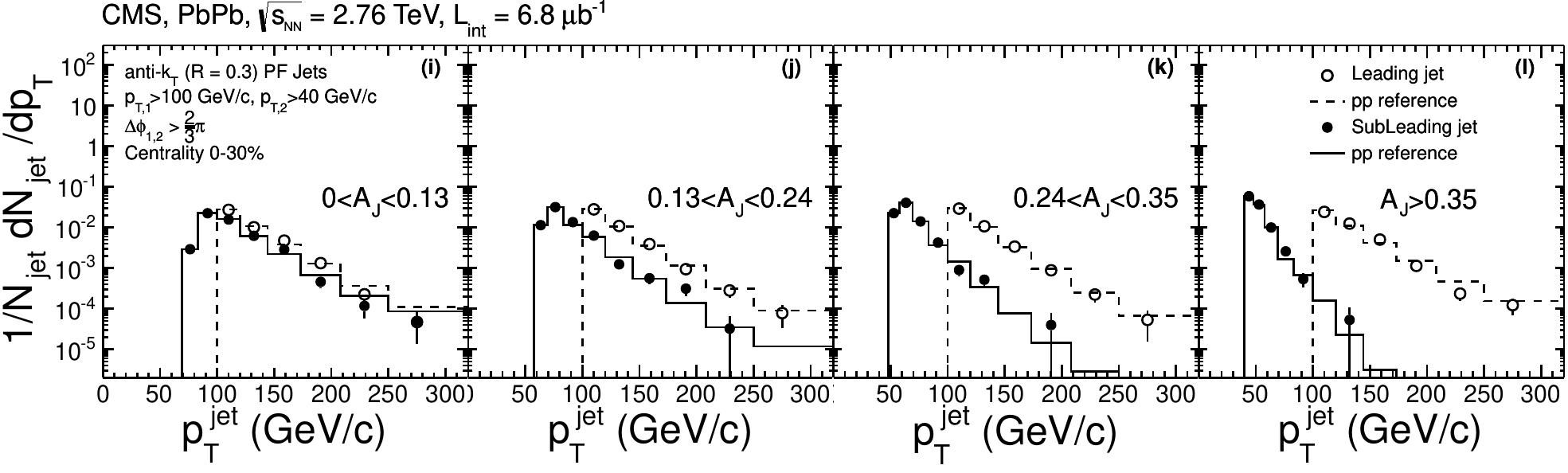}
\end{center}
\caption{(a--d) Fragmentation functions for the leading (open circles) and subleading (solid points) jets in four regions of $\AJ$ in central PbPb collisions compared to the pp reference.
(e--h) Ratio of each fragmentation function to its
pp-based reference.
Error bars shown are statistical. The systematic uncertainty is
represented by hollow boxes (leading jet) or gray boxes (subleading jet).
(i--l) Jet $\pt$ distributions in PbPb collisions in four regions of $\AJ$ (not corrected for efficiency and not unfolded for \pt resolution)
compared to the
pp-based reference
(see text). Only statistical uncertainties are shown.
}
\label{fig:ffLeadSublead_centralAJ_ppDiv}
\end{figure}

To study in more detail the potential effect of medium-induced energy loss on the fragmentation properties of partons, the data sample of central events (0--30\% centrality), where a large average dijet imbalance is observed,
is divided into classes of dijet imbalance.
Four $\AJ$ selections are chosen, which split the central PbPb data sample into
approximately equal number of dijets: $0 \!<\! \AJ \!<\! 0.13$, $0.13 \!<\! \AJ \!<\! 0.24$, $0.24 \!<\! \AJ \!<\! 0.35$, and $\AJ \!>\! 0.35$.
For each of these event classes, the fragmentation functions are constructed separately for the leading and subleading jets.
In Fig.~\ref{fig:ffLeadSublead_centralAJ_ppDiv}(a--d) the fragmentation functions
are shown in bins of increasing dijet imbalance, from left to right. The corresponding
jet \pt distributions used in the fragmentation functions are shown in Fig~\ref{fig:ffLeadSublead_centralAJ_ppDiv}(i--l), illustrating the kinematic range of the measurement and the energy imbalance between leading and subleading jets, as selected by the $\AJ$ interval. The overlaid histogram in the same set of figures shows the pp-based reference distributions.
Ratios of \PbPb\ data to the pp-based reference are shown in Fig.~\ref{fig:ffLeadSublead_centralAJ_ppDiv}(e--h).
In the $\xi$ range of 0.5--4.0 the \PbPb\ and the pp distributions typically agree to within (10--20)\% which is smaller than the systematic uncertainty in the \PbPb\ measurement, as indicated by the size of the shaded area and open boxes for the leading and subleading jet, respectively.
Within uncertainties, the \PbPb\ data and pp-based reference show the same
fragmentation properties, for different jet imbalance in leading, as well as subleading, jets.

\section{Conclusions}
The CMS detector has been used to study jet fragmentation properties
in pp and \PbPb\ collisions at $\sNN=2.76\TeV$ in data
samples corresponding to integrated luminosities of about 231\nbinv
and 6.8\mubinv, respectively.
Jets were reconstructed based on particle-flow objects using the anti-$k_{T}$
sequential clustering algorithm, with a radius parameter of 0.3.
The reconstructed jet momenta were corrected to final-state particle level.
Dijets were selected consisting of a leading jet
with $p_{\mathrm{T},1} > 100$\GeVc and a subleading jet
of $p_{\mathrm{T},2} > 40$~\GeVc, with axes that lie within $|\eta| < 2$.
The azimuthal opening angle $\Delta \phi_{1,2}$ between the leading and subleading jet
was required to be larger than $2\pi/3$.
The selected jets were used to construct the high-$\pt$ component of the
fragmentation functions by correlating their momentum with the momenta of tracks of $\pt > 4\GeVc$ within a cone of $\Delta R < 0.3$ around the jet axis.
The \PbPb\ results were compared to those in a pp-based reference taking into account
the different jet momentum distribution and the effect of fluctuations in the underlying \PbPb\ event on the jet momentum reconstruction.
The jet properties were studied as a function of the collision centrality and the dijet
transverse momentum imbalance.
Central PbPb events show a significant excess of unbalanced jet pairs.
Nevertheless, the fragmentation functions reconstructed in
PbPb collisions for different event centrality and dijet-\pt imbalance
agree within the measurement uncertainty with the pp-based reference
for jets of the same reconstructed momentum.
This shows that, after traversing the dense strongly interacting medium,
partons produced in \PbPb\ collisions are reconstructed as jets with a
significantly reduced momentum. However, the partition of the smaller momentum that remains within the
jet cone into high-\pt particles corresponds to that observed for jets fragmenting in vacuum, as
seen in \pp\ collisions.

\section*{Acknowledgements}
We congratulate our colleagues in the CERN accelerator departments for the excellent performance of the LHC machine. We thank the technical and administrative staff at CERN and other CMS institutes, and acknowledge support from: FMSR (Austria); FNRS and FWO (Belgium); CNPq, CAPES, FAPERJ, and FAPESP (Brazil); MES (Bulgaria); CERN; CAS, MoST, and NSFC (China); COLCIENCIAS (Colombia); MSES (Croatia); RPF (Cyprus); MoER, SF0690030s09 and ERDF (Estonia); Academy of Finland, MEC, and HIP (Finland); CEA and CNRS/IN2P3 (France); BMBF, DFG, and HGF (Germany); GSRT (Greece); OTKA and NKTH (Hungary); DAE and DST (India); IPM (Iran); SFI (Ireland); INFN (Italy); NRF and WCU (Korea); LAS (Lithuania); CINVESTAV, CONACYT, SEP, and UASLP-FAI (Mexico); MSI (New Zealand); PAEC (Pakistan); MSHE and NSC (Poland); FCT (Portugal); JINR (Armenia, Belarus, Georgia, Ukraine, Uzbekistan); MON, RosAtom, RAS and RFBR (Russia); MSTD (Serbia); MICINN and CPAN (Spain); Swiss Funding Agencies (Switzerland); NSC (Taipei); TUBITAK and TAEK (Turkey); STFC (United Kingdom); DOE and NSF (USA).
Individuals have received support from the Marie-Curie programme and the European Research Council (European Union); the Leventis Foundation; the A. P. Sloan Foundation; the Alexander von Humboldt Foundation; the Belgian Federal Science Policy Office; the Fonds pour la Formation \`a la Recherche dans l'Industrie et dans l'Agriculture (FRIA-Belgium); the Agentschap voor Innovatie door Wetenschap en Technologie (IWT-Belgium); the Council of Science and Industrial Research, India; and the HOMING PLUS programme of Foundation for Polish Science, cofinanced from European Union, Regional Development Fund.

\bibliography{auto_generated}   

\providecommand{\href}[2]{#2}\begingroup\raggedright\begin{thebibliography}{10}%
\makeatletter
\providecommand{\hrefCMSnoop }[0]{\@secondoftwo}%
\makeatother
\providecommand{\doi}{\texttt{doi:}\begingroup \urlstyle{tt}\Url}

\bibitem{Bjorken:1982tu}
\href {http://lss.fnal.gov/archive/preprint/fermilab-pub-82-059-t.shtml} {J.~D.
  Bjorken, ``Energy loss of energetic partons in QGP: possible extinction of
  high $p_T$ jets in hadron-hadron collisions'',} FERMILAB-PUB 82-059-THY,
  (1982).

\bibitem{CasalderreySolana:2007zz}
\href {http://th-www.if.uj.edu.pl/acta/vol38/pdf/v38p3731.pdf}
  {J.~Casalderrey-Solana and C.~A. Salgado, ``{Introductory lectures on jet
  quenching in heavy ion collisions}'',} \textit{ Acta Phys. Polon. B} \textbf{
  38} (2007) 3731,
\href{http://www.arXiv.org/abs/0712.3443}{\texttt{ arXiv:0712.3443}}.

\bibitem{d'Enterria:2009am}
\hrefCMSnoop {} {D.~d'Enterria, ``Jet quenching''}, volume 23: Relativistic
  Heavy Ion Physics of \textit{ Springer Materials - The Landolt-B{\"or}nstein
  Database}, ch.~6.4.
\newblock Springer-Verlag, 2010.
\newblock \href{http://www.arXiv.org/abs/0902.2011}{\texttt{ arXiv:0902.2011}}.
\newblock
\href{http://dx.doi.org/10.1007/978-3-642-01539-7_16}{\doi{10.1007/978-3-642-01539-7_16}}.

\bibitem{Arsene:2004fa}
\hrefCMSnoop {} {{ BRAHMS} Collaboration, ``{Quark Gluon Plasma and Color Glass
  Condensate at RHIC? The perspective from the BRAHMS experiment}'',} \textit{
  Nucl. Phys. A} \textbf{ 757} (2005) 1,
  \href{http://dx.doi.org/10.1016/j.nuclphysa.2005.02.130}{\doi{10.1016/j.nuclphysa.2005.02.130}},
\href{http://www.arXiv.org/abs/nucl-ex/0410020}{\texttt{
  arXiv:nucl-ex/0410020}}.

\bibitem{Back:2004je}
\hrefCMSnoop {} {{ PHOBOS} Collaboration, ``{The PHOBOS perspective on
  discoveries at RHIC}'',} \textit{ Nucl. Phys. A} \textbf{ 757} (2005) 28,
  \href{http://dx.doi.org/10.1016/j.nuclphysa.2005.03.084}{\doi{10.1016/j.nuclphysa.2005.03.084}},
\href{http://www.arXiv.org/abs/nucl-ex/0410022}{\texttt{
  arXiv:nucl-ex/0410022}}.

\bibitem{Adams:2005dq}
\hrefCMSnoop {} {{ STAR} Collaboration, ``{Experimental and theoretical
  challenges in the search for the quark gluon plasma: The STAR collaboration's
  critical assessment of the evidence from RHIC collisions}'',} \textit{ Nucl.
  Phys. A} \textbf{ 757} (2005) 102,
  \href{http://dx.doi.org/10.1016/j.nuclphysa.2005.03.085}{\doi{10.1016/j.nuclphysa.2005.03.085}},
\href{http://www.arXiv.org/abs/nucl-ex/0501009}{\texttt{
  arXiv:nucl-ex/0501009}}.

\bibitem{Adcox:2004mh}
\hrefCMSnoop {} {{ PHENIX} Collaboration, ``{Formation of dense partonic matter
  in relativistic nucleus nucleus collisions at RHIC: Experimental evaluation
  by the PHENIX collaboration}'',} \textit{ Nucl. Phys. A} \textbf{ 757} (2005)
  184,
  \href{http://dx.doi.org/10.1016/j.nuclphysa.2005.03.086}{\doi{10.1016/j.nuclphysa.2005.03.086}},
\href{http://www.arXiv.org/abs/nucl-ex/0410003}{\texttt{
  arXiv:nucl-ex/0410003}}.

\bibitem{Aamodt:2010jd}
\hrefCMSnoop {} {{ ALICE} Collaboration, ``{Suppression of Charged Particle
  Production at Large Transverse Momentum in Central Pb--Pb Collisions at
  $\sqrt{s_{NN}} = 2.76$ TeV}'',} \textit{ Phys. Lett. B} \textbf{ 696} (2011)
  30,
  \href{http://dx.doi.org/10.1016/j.physletb.2010.12.020}{\doi{10.1016/j.physletb.2010.12.020}},
\href{http://www.arXiv.org/abs/1012.1004}{\texttt{ arXiv:1012.1004}}.

\bibitem{CMS-HIN-10-005}
\hrefCMSnoop {} {{ CMS} Collaboration, ``Study of high-$p_T$ charged particle
  suppression in PbPb compared to pp collisions at $\sqrt{s_{NN}} =
  2.76\:\mathrm{TeV}$'',} \textit{ Eur. Phys. J. C} \textbf{ 72} (2012) 1945,
  \href{http://dx.doi.org/10.1140/epjc/s10052-012-1945-x}{\doi{10.1140/epjc/s10052-012-1945-x}},
  \href{http://www.arXiv.org/abs/1202.2554}{\texttt{ arXiv:1202.2554}}.

\bibitem{Aad:2012bu}
\hrefCMSnoop {} {{ ATLAS} Collaboration, ``{Measurement of the azimuthal
  anisotropy for charged particle production in $\sqrt{s_{NN}}$ = 2.76 TeV
  lead-lead collisions with the ATLAS detector}'',} (2012).
  \href{http://www.arXiv.org/abs/1203.3087}{\texttt{ arXiv:1203.3087}}.
Submitted to Phys. Rev. C.

\bibitem{Chatrchyan:2012xq}
\hrefCMSnoop {} {{ CMS} Collaboration, ``{Azimuthal anisotropy of charged
  particles at high transverse momenta in PbPb collisions at $\sqrt{s_{NN}}$ =
  2.76 TeV}'',} (2012). \href{http://www.arXiv.org/abs/1204.1850}{\texttt{
  arXiv:1204.1850}}.
Submitted to \textit{Phys. Rev. Lett.}

\bibitem{Aamodt:2010pa}
\hrefCMSnoop {} {{ ALICE} Collaboration, ``{Elliptic flow of charged particles
  in Pb-Pb collisions at 2.76 TeV}'',} \textit{ Phys. Rev. Lett.} \textbf{ 105}
  (2010) 252302,
  \href{http://dx.doi.org/10.1103/PhysRevLett.105.252302}{\doi{10.1103/PhysRevLett.105.252302}},
\href{http://www.arXiv.org/abs/1011.3914}{\texttt{ arXiv:1011.3914}}.

\bibitem{Chatrchyan:2012ta}
\hrefCMSnoop {} {{ CMS} Collaboration, ``{Measurement of the elliptic
  anisotropy of charged particles produced in PbPb collisions at
  nucleon-nucleon center-of-mass energy = 2.76 TeV}'',} (2012).
  \href{http://www.arXiv.org/abs/1204.1409}{\texttt{ arXiv:1204.1409}}.
Submitted to \textit{Phys. Rev. C}.

\bibitem{ATLAS:2011ah}
\hrefCMSnoop {} {{ ATLAS} Collaboration, ``{Measurement of the pseudorapidity
  and transverse momentum dependence of the elliptic flow of charged particles
  in lead-lead collisions at $\sqrt{s_{NN}}$ = 2.76 TeV with the ATLAS
  detector}'',} \textit{ Phys. Lett. B} \textbf{ 707} (2012) 330,
  \href{http://dx.doi.org/10.1016/j.physletb.2011.12.056}{\doi{10.1016/j.physletb.2011.12.056}},
\href{http://www.arXiv.org/abs/1108.6018}{\texttt{ arXiv:1108.6018}}.

\bibitem{Collaboration:2010bu}
\hrefCMSnoop {} {{ ATLAS} Collaboration, ``{Observation of a
  Centrality-Dependent Dijet Asymmetry in Lead-Lead Collisions at
  $\sqrt{s_{NN}}$~= 2.76 TeV with the ATLAS Detector at the LHC}'',} \textit{
  Phys. Rev. Lett.} \textbf{ 105} (2010) 252303,
  \href{http://dx.doi.org/10.1103/PhysRevLett.105.252303}{\doi{10.1103/PhysRevLett.105.252303}},
  \href{http://www.arXiv.org/abs/1011.6182}{\texttt{ arXiv:1011.6182}}.

\bibitem{Chatrchyan:2011sx}
\hrefCMSnoop {} {{ CMS} Collaboration, ``{Observation and studies of jet
  quenching in PbPb collisions at nucleon-nucleon center-of-mass energy = 2.76
  TeV}'',} \textit{ Phys. Rev. C} \textbf{ 84} (2011) 024906,
  \href{http://dx.doi.org/10.1103/PhysRevC.84.024906}{\doi{10.1103/PhysRevC.84.024906}},
\href{http://www.arXiv.org/abs/1102.1957}{\texttt{ arXiv:1102.1957}}.

\bibitem{CMS-HIN-11-013}
\hrefCMSnoop {} {{ CMS} Collaboration, ``{Jet momentum dependence of jet
  quenching in PbPb collisions at $\sqrt{s_{NN}}$ = 2.76 TeV}'',} \textit{
  Phys. Lett. B} \textbf{ 712} (2012) 176,
  \href{http://dx.doi.org/10.1016/j.physletb.2012.04.058}{\doi{10.1016/j.physletb.2012.04.058}},
\href{http://www.arXiv.org/abs/1202.5022}{\texttt{ arXiv:1202.5022}}.

\bibitem{Dokshitzer:1991wu}
Y.~L. Dokshitzer and V.~A. Khoze, ``{Basics of perturbative QCD}''.
\newblock {Editions Fronti\`eres},
1991.
\newblock

\bibitem{Guo:2000nz}
\hrefCMSnoop {} {X.-F. Guo and X.-N. Wang, ``{Multiple scattering, parton
  energy loss and modified fragmentation functions in deeply inelastic e A
  scattering}'',} \textit{ Phys. Rev. Lett.} \textbf{ 85} (2000) 3591,
  \href{http://dx.doi.org/10.1103/PhysRevLett.85.3591}{\doi{10.1103/PhysRevLett.85.3591}},
\href{http://www.arXiv.org/abs/hep-ph/0005044}{\texttt{ arXiv:hep-ph/0005044}}.

\bibitem{Borghini:2005em}
\hrefCMSnoop {} {N.~Borghini and U.~A. Wiedemann, ``{Distorting the hump-backed
  plateau of jets with dense QCD matter}'',} (2005).
  \href{http://www.arXiv.org/abs/hep-ph/0506218}{\texttt{
  arXiv:hep-ph/0506218}}.
CERN-PH-TH-2005-100, BI-TP-2005-20.

\bibitem{Armesto:2007dt}
\hrefCMSnoop {} {N.~Armesto {et~al.}, ``{Medium-evolved fragmentation
  functions}'',} \textit{ JHEP} \textbf{ 02} (2008) 048,
  \href{http://dx.doi.org/10.1088/1126-6708/2008/02/048}{\doi{10.1088/1126-6708/2008/02/048}},
\href{http://www.arXiv.org/abs/0710.3073}{\texttt{ arXiv:0710.3073}}.

\bibitem{Arleo:2008dn}
\hrefCMSnoop {} {F.~Arleo, ``{(Medium-modified) Fragmentation Functions}'',}
  \textit{ Eur. Phys. J. C} \textbf{ 61} (2009) 603,
  \href{http://dx.doi.org/10.1140/epjc/s10052-009-0871-z}{\doi{10.1140/epjc/s10052-009-0871-z}},
\href{http://www.arXiv.org/abs/0810.1193}{\texttt{ arXiv:0810.1193}}.

\bibitem{Cacciari:2008gp}
\hrefCMSnoop {} {M.~Cacciari, G.~P. Salam, and G.~Soyez, ``{The anti-$k_t$ jet
  clustering algorithm}'',} \textit{ JHEP} \textbf{ 04} (2008) 063,
  \href{http://dx.doi.org/10.1088/1126-6708/2008/04/063}{\doi{10.1088/1126-6708/2008/04/063}},
\href{http://www.arXiv.org/abs/0802.1189}{\texttt{ arXiv:0802.1189}}.

\bibitem{bib_CMS}
\hrefCMSnoop {} {{ CMS} Collaboration, ``The {CMS} experiment at the {CERN}
  {LHC}'',} \textit{ JINST} \textbf{ 3} (2008) S08004,
\href{http://dx.doi.org/10.1088/1748-0221/3/08/S08004}{\doi{10.1088/1748-0221/3/08/S08004}}.

\bibitem{MattPFlow}
\hrefCMSnoop {} {M.~Nguyen {et~al.}, ``{Jet Reconstruction with Particle Flow
  in Heavy-Ion Collisions with CMS}'',} \textit{ J. Phys. G} \textbf{ 38}
  (2011) 124151,
  \href{http://dx.doi.org/10.1088/0954-3899/38/12/124151}{\doi{10.1088/0954-3899/38/12/124151}},
\href{http://www.arXiv.org/abs/1107.0179}{\texttt{ arXiv:1107.0179}}.

\bibitem{particle_flow}
\href {http://cdsweb.cern.ch/record/1194487} {{ CMS} Collaboration,
  ``Particle--Flow Event Reconstruction in {CMS} and Performance for Jets,
  Taus, and {\MET}'',} CMS Physics Analysis Summary CMS-PAS-PFT-09-001, (2009).

\bibitem{Kodolova:2007hd}
O.~Kodolova\hrefCMSnoop {} { {et~al.}, ``{The performance of the jet
  identification and reconstruction in heavy ions collisions with CMS
  detector}'',} \textit{ Eur. Phys. J. C} \textbf{ 50} (2007) 117,
\href{http://dx.doi.org/10.1140/epjc/s10052-007-0223-9}{\doi{10.1140/epjc/s10052-007-0223-9}}.

\bibitem{bib_pythia}
\hrefCMSnoop {} {T.~Sj{\"o}strand, S.~Mrenna, and P.~Skands, ``{PYTHIA} 6.4
  physics and manual'',} \textit{ JHEP} \textbf{ 05} (2006) 026,
  \href{http://dx.doi.org/10.1088/1126-6708/2006/05/026}{\doi{10.1088/1126-6708/2006/05/026}},
  \href{http://www.arXiv.org/abs/hep-ph/0603175}{\texttt{
  arXiv:hep-ph/0603175}}. Tune D6T with CTEQ6L1 PDFs.

\bibitem{TUNE-D6T-1}
\href {http://th-www.if.uj.edu.pl/acta/vol39/pdf/v39p2611.pdf} {R.~Field,
  ``Physics at the Tevatron'',} \textit{ Acta Phys. Polon. B} \textbf{ 39}
  (2008) 2611.

\bibitem{TUNE-D6T-2}
\href {http://www-library.desy.de/preparch/desy/proc/proc09-06.pdf}
  {P.~Bartalini and L.~Fan{\'o}, eds., ``{Proceedings of the First
  International Workshop on Multiple Partonic Interactions at the LHC MPI'08,
  October 27-31, 2008}''}.
\newblock Perugia, Italy, (2009).
\newblock
\href{http://www.arXiv.org/abs/1003.4220}{\texttt{ arXiv:1003.4220}}.

\bibitem{CTEQ6L1}
J.~Pumplin\hrefCMSnoop {} { {et~al.}, ``{New generation of parton distributions
  with uncertainties from global QCD analysis}'',} \textit{ JHEP} \textbf{
  0207} (2002) 012,
  \href{http://dx.doi.org/10.1088/1126-6708/2002/07/012}{\doi{10.1088/1126-6708/2002/07/012}},
\href{http://www.arXiv.org/abs/hep-ph/0201195}{\texttt{ arXiv:hep-ph/0201195}}.

\bibitem{Chatrchyan:2011ds}
\hrefCMSnoop {} {{ CMS} Collaboration, ``{Determination of Jet Energy
  Calibration and Transverse Momentum Resolution in CMS}'',} \textit{ J.
  Instrum.} \textbf{ 06} (2011) P11002,
  \href{http://dx.doi.org/10.1088/1748-0221/6/11/P11002}{\doi{10.1088/1748-0221/6/11/P11002}},
\href{http://www.arXiv.org/abs/1107.4277}{\texttt{ arXiv:1107.4277}}.

\bibitem{bib_geant}
\hrefCMSnoop {} {S.~Agostinelli {et~al.}, ``{Geant4}---a simulation toolkit'',}
  \textit{ Nucl. Instrum. Meth. A} \textbf{ 506} (2003) 250,
  \href{http://dx.doi.org/10.1016/S0168-9002(03)01368-8}{\doi{10.1016/S0168-9002(03)01368-8}}.

\bibitem{Lokhtin:2005px}
\hrefCMSnoop {} {I.~P. Lokhtin and A.~M. Snigirev, ``{A model of jet quenching
  in ultrarelativistic heavy ion collisions and high-p$_{T}$ hadron spectra at
  RHIC}'',} \textit{ Eur. Phys. J. C} \textbf{ 45} (2006) 211,
  \href{http://dx.doi.org/10.1140/epjc/s2005-02426-3}{\doi{10.1140/epjc/s2005-02426-3}},
\href{http://www.arXiv.org/abs/hep-ph/0506189}{\texttt{ arXiv:hep-ph/0506189}}.

\bibitem{PhysRevD.64.032001}
\hrefCMSnoop {} {{ CDF} Collaboration, ``Measurement of the inclusive jet cross
  section in $p\bar{p}$ collisions at $\sqrt{s}=1.8$ TeV'',} \textit{ Phys.
  Rev. D} \textbf{ 64} (2001) 032001,
  \href{http://dx.doi.org/10.1103/PhysRevD.64.032001}{\doi{10.1103/PhysRevD.64.032001}}.

\bibitem{PhysRevLett.65.968}
\hrefCMSnoop {} {F.~Abe {et~al.}, ``Jet-fragmentation properties in $p\bar{p}$
  collisions at $\sqrt{s}=1.8$ TeV'',} \textit{ Phys. Rev. Lett.} \textbf{ 65}
  (1990) 968,
  \href{http://dx.doi.org/10.1103/PhysRevLett.65.968}{\doi{10.1103/PhysRevLett.65.968}}.

\end{thebibliography}\endgroup

\cleardoublepage \appendix\section{The CMS Collaboration \label{app:collab}}\begin{sloppypar}\hyphenpenalty=5000\widowpenalty=500\clubpenalty=5000\textbf{Yerevan Physics Institute,  Yerevan,  Armenia}\\*[0pt]
S.~Chatrchyan, V.~Khachatryan, A.M.~Sirunyan, A.~Tumasyan
\vskip\cmsinstskip
\textbf{Institut f\"{u}r Hochenergiephysik der OeAW,  Wien,  Austria}\\*[0pt]
W.~Adam, T.~Bergauer, M.~Dragicevic, J.~Er\"{o}, C.~Fabjan\cmsAuthorMark{1}, M.~Friedl, R.~Fr\"{u}hwirth\cmsAuthorMark{1}, V.M.~Ghete, J.~Hammer, N.~H\"{o}rmann, J.~Hrubec, M.~Jeitler\cmsAuthorMark{1}, W.~Kiesenhofer, V.~Kn\"{u}nz, M.~Krammer\cmsAuthorMark{1}, D.~Liko, I.~Mikulec, M.~Pernicka$^{\textrm{\dag}}$, B.~Rahbaran, C.~Rohringer, H.~Rohringer, R.~Sch\"{o}fbeck, J.~Strauss, A.~Taurok, P.~Wagner, W.~Waltenberger, G.~Walzel, E.~Widl, C.-E.~Wulz\cmsAuthorMark{1}
% ...
\vskip\cmsinstskip
\dag:~Deceased\\
1:~~Also at Vienna University of Technology, Vienna, Austria\\
% ...

\end{sloppypar}
\end{document}